# Modeling pattern formation in communities by using information particles

Junichi Miyakoshi[1]

[1]Hitachi, Ltd. Research & Development Group, Kyoto University, International Science Innovation Building 405, Yoshida-honmachi, Sakyo-ku, Kyoto-shi, Kyoto, 606-8501 Japan.

**Understanding the pattern formation in communities has been at the center of attention in various fields. Here we introduce a novel model, called an "information-particle model," which is based on the reaction-diffusion model and the distributed behavior model. The information particle drives competition or coordination among species. Therefore, a traverse of information particles in a social system makes it possible to express four different classes of patterns (i.e. "stationary", "competitive-equilibrium", "chaotic", and "periodic"). Remarkably, "competitive equilibrium" well expresses the complex dynamics that is equilibrium macroscopically and non-equilibrium microscopically. Although it is a fundamental phenomenon in pattern formation in nature, it has not been obtained by conventional models. Furthermore, the pattern transitions across the classes depending only on parameters of system, namely, the number of species (vertices in network) and distance (edges) between species. It means that one information-particle model successfully develops the patterns with an in-situ computation under various environments.**

Pattern formation in communities is one of the classical challenges to unveil coexistence of different species or metapopululation dynamics in nature[1, 2, 3], so it firmly constitutes a research field. In the case of phenomenon, the species compete and coordinate with each other, and communities are organized with various patterns. Natural communities are characterized by two features: formation of spatial-heterogeneous patterns and dynamic systems[4]. As for the first, complex patterns, identified by spatial discontinuities in the distribution of species, are formed. As for the second, local densities of species, as well as the relative abundances of species, change with time. For instance, in planktonic communities or in tropical rain forests, a large number of species coexist in a macroscopic homogeneous habitat with little scope for spatial or temporal niche differentiation[3, 5]. Namely, natural communities reach a competitive equilibrium. In the past decade, attention has turned to pattern formation on social networks. The communities are organized autonomously according to social activities, such as exchanging messages[6, 7, 8, 9], sharing blogs[10], and others[11, 12].

To study the above-described phenomenon, namely, pattern formation in communities, a reaction-diffusion model has been introduced. The Turing model[13], devised by Alan Turing at 1952, is one of the best-known models of pattern formation. Turing showed that differences in diffusion rates may destabilize the uniform state of reacting species and that a combination of known physical elements is sufficient to explain biological pattern formation. The model has been extensively studied in the fields of biological[14-19], chemical[20-23] and physical systems[24], fundamental analysis[25-28], and social systems such as distributed ecological systems and human communities[29-32]. A social system is expressed as a network by discretizing continuous space. The pray-predator model, also known as the Lotka-Volterra competition model, has been used to study pattern formation in communities[3, 33, 34, 35]. The model can form various patterns, such as stationary, periodic, and chaotic. On the other hand, the dynamics of pattern formation has also been studied from the perspective of the distributed-behavior model[36]. In this model, the components of a community, such as species, are expressed as particles. The aggregate motion of a flock of a species (e.g., birds) is elaborated by the dense interaction of the relatively simple behaviors of the individual simulated species. The model has been applied to control of a robot flock system[37, 38], namely, flock-formation control and collision avoidance. It can thus express various dynamics of a flock, herd or community.

The two models, namely, the reaction-diffusion model and the distributed-behavior model, feature either various patterns or various dynamics, although pattern formation in natural communities has both of these features. In particular, a dynamics, which is called the competitive equilibrium, has not been expressed by the conventional models. In this article, we investigate pattern formation with a model based on both models. This model expresses species as particles, defined as "information particles," in the same manner as the distributed-behavior model. The information particles diffuse in a network-structured space in the same way as species expressed by the reaction-diffusion model. The model can thus form various patterns and simulate complex dynamics.

## Results

**Reaction-diffusion model.** The reaction-diffusion model is described by the following differential equation:

$$\begin{cases} \frac{\partial}{\partial t} u(\boldsymbol{x},t) = f(u,v) + D_{act} \nabla^2 u(\boldsymbol{x},t) \\ \frac{\partial}{\partial t} v(\boldsymbol{x},t) = g(u,v) + D_{inh} \nabla^2 \mathrm{v}(\boldsymbol{x},t) \end{cases} \quad (1)$$

where $u(\boldsymbol{x},t)$ and $v(\boldsymbol{x},t)$ are respectively local densities of activator and inhibitor species. Function $f(u,t)$ specifies dynamics of the activator that performs self-enhancement of a production of the activator. Function $\mathrm{g}(u,t)$ specifies dynamics of the inhibitor, which performs an antagonistic reaction against growth of the activator. $D_{act}$ and $D_{inh}$ are respectively the diffusion constants of the activator and inhibitor. An increase in ratio $D_{inh}/D_{act}$ leads to spontaneous development of self-organized patterns, which are obtained by solving Eq. (1). The model can form various patterns (see Methods for their details).

**Information-particle model and algorithm.** The proposed model for evolving pattern formation in communities is based on the reaction-diffusion model and the distributed-behavior model. In this combined model, a species is expressed as a particle, defined as an "information particle." These particles traverse in a network-structured space; namely, they can move between two nodes along an edge connecting the nodes. It is considered that the network is undirected and an edge is weighted. The weight represents distance between nodes.

The particle has information to identify the type of particle. The type corresponds to the kind of species, such as the activator and inhibitor of the reaction-diffusion model. Density $u_i(t)$ in Eq. (1) can be expressed by the number of particles on a node. It is considered that the information-particle model includes two states, $\{u, v\}$. The state on node $i$ is obtained from the number of particles in state $u$, $N_i^u$, and the number of particles in state $v$, $N_i^v$. The state of each node changes according to $N_i^u$ and $N_i^v$. The information particle on a node can transfer to the connected node along an edge; namely, particles traverse a network-structured space. The transfer criterion governing a particle is given by

$$\begin{cases} Tr(u_i) = f(N_i^u, N_i^v) + \epsilon \sum_{j=1}^{N} \frac{A_{ij}}{D_j} (N_i^u - \sigma N_i^v) \\ Tr(v_i) = f(N_i^u, N_i^v) + \epsilon \sum_{j=1}^{N} \frac{A_{ij}}{D_j} (N_i^v - \sigma N_i^u) \end{cases} \quad (2)$$

for $i = 1,2,\cdots,N$. Here, $Tr(u_i)$ and $Tr(v_i)$ represent a transfer criterion for node $i$, and $D_j$ represents the number of edges, namely, a degree, of node *j*. The transfer criterion for an information particle on a node depends on the number of particles on the same node and the number of particles on neighboring nodes. The information particle tends to transfer to a neighboring node if the number of information particles in a different state is more than that of information particles in the same state as the particle in question; otherwise, the information particle tends not to transfer. This transfer behavior corresponds to competition and coordination observed in pattern formation in communities in nature.

A particle transfers between nodes at a certain velocity, and it reaches its destination node in a finite time. That is, the travel time represents a time delay from the time the particle departs the source node until it reaches the destination node. The influence of such a time delay in the case of pattern formation has been studied[39]. A pattern is organized as a result of the flow of particles on a network. The pattern is specified by the distribution of the particles. The state of a node is determined by the number of particles on the node. For example, when the number of states of particles is two, $\{u, v\}$, the number of states of a node is also two. If $N_i^u > N_i^v$ at node *i*, the state of node *i* is $u$; otherwise, it is $v$.

An algorithm for evaluating the information-particle model is proposed as follows. The algorithm is comprised of three phases, an initial phase, a traverse phase, and an evaluation phase, which operate in order as follows:

1. In the initial phase, a particle is distributed to each node.
2. In the traverse phase, in consecutive time steps, $t = 0,1,\cdots,T$, a particle traverses the network along the edges between nodes. The steps of the traverse process are described below.

   Step 1. The transfer criterion (*Tr*) is calculated, and if *Tr* > 0, step 2 is executed. Otherwise, time *t* is increased, and step 1 is re-executed.

   Step 2. Destination node *j* is randomly selected from the neighborhood of node *i*. A particle travels from node *i* to node *j* at velocity. Step 3 is then executed.

   Step 3. If the particle reaches node *j*, step 1 is returned to. Otherwise, time *t* is increased, and step 3 is re-executed.

3. In the evaluation phase, the particle distributions of all nodes are evaluated. The pattern of the entire network is obtained.

**Results of evaluation of information-particle model.** Pattern formation was examined by using the information-particle model. The criteria used to quantify the pattern formation, namely, the mutual information, $I_{ave}$ and $I_{dif}$, the amplitude of pattern, $\overline{A}$, and the power spectrum of the amplitude, $PW_{AC}$, are considered in the following (see Methods for their details). A regular-grid network was used in this evaluation. To analyze the pattern formation, parameter $r$ is defined as the ratio of the number of particles to the size of the entire network:

$$r = \frac{N_{particle}}{N_{node}} \tag{3}$$

where $N_{particle}$ is the total number of particles, and $N_{node}$ is the total number of nodes in the entire network. According to Eq. (3), $r$ represents the density (or population) of particles. Traverse criteria $Tr$ depends on the number of particles, so $r$ affects $Tr$ in the same manner as parameter $\lambda$ [40].

Patterns revealed by this model as a function of $r$ are shown in Fig. 1. The figure shows that the patterns change dramatically with changing $r$ and can be categorized into four classes according to criteria, $I_{ave}$, $I_{dif}$, and $PW_{AC}$. The values of the criteria are obtained after enough time steps (two million) to reach a stable pattern. The classes are defined as corresponding to the classes given by Wolfram[42]. In the first class ("class 1" (stationary) hereafter), $I_{ave}$, $I_{dif}$, and $PW_{AC}$ are zero. In the second class 2 ("class 2" (competitive-equilibrium)), $I_{dif}$ has a peak, and the other criteria are low. In the third class ("class 3" (chaotic)), $I_{ave}$ is high, and the other criteria are low. In the fourth ("class 4" (periodic)), $I_{ave}$ and $PW_{AC}$ are high. In particular, $PW_{AC}$ has a peak. Patterns and dynamics in the four classes are investigated below.

**Class 1.** In class 1, $I_{ave}$, $I_{dif}$ and $PW_{AC}$, are zero. This condition indicates that the pattern does not change temporally; namely, all particles are stationary, and the pattern is stationary. On the other hand, the pattern is spatially complex. Figure 2(a) illustrates the temporal changes of amplitude, $\overline{A}$. In the initial period (t < 10,000 steps), $\overline{A}$ varies greatly. After that period, $\overline{A}$ converges. In the zoomed graph of Fig. 2(a) (right panel), $\overline{A}$ is constant. The temporal trajectory of $\overline{A}$ is shown in Figure 3(a). The trajectory is a point that does not move. In other words, the pattern is fixed temporally. Figure 3(e) shows a map that displays the spatial distribution of the mutual information ($I_{ave}$ and $I_{dif}$). The map indicates that the mutual information in the entire network is low and that the pattern is spatially and temporally stationary.

**Class 2.** In class 2, $I_{dif}$ has a peak, and the other criteria, $I_{ave}$ and $PW_{AC}$, are not zero. The temporal behavior of the pattern is analyzed first. The dynamics of $\overline{A}$ is shown in Fig. 2(b), which shows the dynamics change little and at random after the initial period (t < 10,000 steps). The trajectory of the amplitude is shown in Fig. 3(b). It is also random in the same way as the dynamics. According to this analysis of the dynamics and trajectory, changes of the pattern do not have a particular cycle. Spatial behavior of the pattern is analyzed next. High $I_{dif}$, as shown in Fig. 1, indicates that a spatial-complex pattern is formed. Figure 4(b) shows a map of the mutual information. A white vertical line is observed at the center of the map. As shown in Fig. 1(c), the line is located at the same position as a boundary between different states, namely, a boundary between red and green regions. Regions excluding the boundary are low mutual information (other regions are black). In other words, the boundary has only dynamics, and other regions are stationary. The dynamics of the pattern formation is described in detail as follows. Figure 4 shows snapshots of the pattern at different time steps. At $t$ = 0 (Fig. 4(a)), no pattern exists. In this case, in the initial phase, all particles are located at the center of the network. At $t$ = 1,000 (Fig. 4(b)), the particles diffuse from the center. At $t$ = 3,000 (Fig. 4(c)), a complicated pattern begins to emerge partially as the particles are distributed to the whole network. At $t$ = 10,000 (Fig. 4(d)), the pattern has a complicated structure. From $t$ = 50,000 to 400,000 (Figs. 4(d) to 4(i)), the boundary in the pattern changes gradually. After $t$ = 400,000, the pattern becomes almost stationary, and only the boundary fluctuates. The dynamics from $t$ = 10,000 to 400,000 indicates that the boundary is autonomously optimized by competition between different particles; therefore, the length of the boundary is minimized. After t = 400,000 (Fig. 4(j)), the dynamics of the pattern becomes macroscopically stable. In contrast, the microscopic dynamics of the boundary continues to change; namely, the particles continuously traverse the network. This dynamics is equal to dynamics of competitive equilibrium. Accordingly, in class 2, the pattern becomes macroscopically stationary and microscopically dynamic. The model thus expresses a pattern that reaches competitive-equilibrium state.

**Class 3.** In class 3, $I_{ave}$ is relatively high, but the other criteria, $I_{dif}$ and $PW_{AC}$ are low. Since $I_{ave}$ does not have a peak, the pattern is not structured. Figures 2(d) shows temporal changes of amplitude, and Figure 3(d) shows a trajectory of the amplitude. The figures show that the amplitude changes randomly, so the particles do not stop every time step. Since $PW_{AC}$ is also low, the pattern does not have a particular cycle. A map of mutual information in Fig. 3(h) shows that the pattern is uniform spatially. The spatial and temporal changes of the pattern are thus chaotic.

**Class 4** In class 4, $I_{ave}$ and $PW_{AC}$ are high. The temporal behavior of the pattern is analyzed as follows. Temporal changes of the amplitude are shown in Fig. 2(c). It shows the amplitude oscillates. A trajectory of the amplitude is shown in Fig. 3(c). The trajectory is periodic, but it fluctuates every period. This behavior, called a strange attractor, is also observed in pattern formation by using the reaction-diffusion model. The power spectrum, $PW_{AC}$, is displayed in Figs. 3(i)-3(l). In Fig. 3(k), a peak of the power spectrum occurs at a certain time step; that is, the pattern periodically changes in 18 time steps. The other classes do not have a peak of $PW_{AC}$, so they are not periodic. In this evaluation, the velocity of a particle is taken as 1.0. The travel time between nodes is 16 time steps and the stationary time at a node is two time steps. A total travel time steps that is a sum of the travel time and the stationary time is 18. The total time steps equals a period of the pattern change. The period of change of the pattern in class 4 depends on the total travel time or the velocity of the particle.

The spatial behavior of the pattern is analyzed as follows. A map of mutual information in Fig. 3(g) shows a non-uniform pattern. It indicates the pattern is structured. Some snapshots of the pattern are shown in Fig. 5. Figure 5(a) displays the trajectory in one period (18 time steps), and Figs. 5(b) to 5(f) show the corresponding patterns. The pattern changes from Figs. 5(b) to 5(f). The pattern in Fig. 5(b) is relatively class-2-like. The patterns in Figs. 5(c) to 5(e) are chaotic (like those of class 3). After the pattern in Fig. 5(e) is formed, the pattern returns to class-2-like in Fig. 5(f). The dynamics of the pattern therefore becomes a periodic state oscillating between the structured pattern (class-2-like) and the chaotic pattern (class-3-like).

**Analysis of classes.** Various patterns spontaneously emerge when the information-particle model is used. The patterns can be classified into four classes by the following criteria: mutual information, $I_{ave}$ and $I_{dif}$, amplitude $\overline{\mathrm{A}}$, and the power $PW_{AC}$. These classes represent the stationary, the competitive-equilibrium, the periodic, and the chaotic pattern. An algorithm for classifying the pattern is defined as a relation between these criteria (see "Method").

Figure 6 illustrates the class diagram that is classified by the algorithm. The parameters in Fig. 6 are $r$ (namely, the ratio of the number of particles to the size of the entire network) and the travel time (or velocity of a particle). In Fig. 6, with increasing $r$, the class changes in the order of class 1, class 2, class 4, and class 3. On the other hand, by increasing the velocity (shortening the travel time), regions of class 2 and class 4 become gradually narrower. When the velocity exceeds a threshold, these classes disappear. Consequently, it is clear that $r$ and the velocity are important factors in pattern formations. Hysteresis is also observed in the same manner as other non-linear systems. It is seen in the class changed by increasing or decreasing $r$ (see "Supplementary materials").

## Discussion

The two models, a reaction-diffusion model and a distributed-behavior model, are introduced to understand the principle of pattern formation. The reaction-diffusion model provides a paradigm of non-equilibrium pattern formation. It has been applied for clarifying pattern formation in biological systems. Recently, it was extensively applied to a network system. This model can create complex patterns. The distributed-behavior model expresses components of a community as particles. The complex dynamics of a whole system is elaborated by interaction of the particles. However, although pattern formation in communities features both formation of complex patterns and complex dynamics, a model that has both of these features has not been reported. In particular, a dynamics, which is called the competitive equilibrium, has not been expressed by the conventional models.

A new model, called an information-particle model, is introduced for understanding pattern formation. This model expresses species as particles, defined as information particles, in the same manner as the distributed-behavior model. The information particles diffuse in a network-structured space in the same way as a species expressed by the reaction-diffusion model. The information-particle model can take advantage of these two models. The above-described evaluation of the model shows that various patterns and complex dynamics emerge when the information-particle model is used. The patterns can be categorized as four classes according to their behavior. In particular, the second class behaves differently from the corresponding class in the classical model. The dynamics of a pattern in the second class is stable macroscopically, and its microscopic behavior constantly changes even if enough time steps elapse. The dynamics is the same as the competitive equilibrium observed in pattern formation of a natural community. Consequently, patterns expressed by using the information-particle model feature both formation of complex patterns and complex dynamics, so the dynamics of pattern formation can be simulated by the model. The model will thus play a fundamental role in understanding pattern formation of communities.

## Methods

**Network-organized reaction-diffusion model.** A reaction-diffusion model has been applied to network-organized systems[31]. In this model, continuous space is expressed by a network that consists of discrete nodes and edges connecting nodes. For instance, activator and inhibitor species diffuse to nodes along an edge. A network is considered to consist of $N$ nodes labeled by indices $i = 1,2,\cdots,N$. A network Laplacian matrix, $L$, has elements $L_{ij} = A_{ij} - \sum_{j=1}^{N} A_{ij}\delta_{ij}$, where $A$ is the adjacency matrix, and elements $A_{ij}$ take values of 1 if node $i$ is connected to node $j$ $(i \neq j)$; otherwise, $A_{ij} = 0$. The network-organized reaction-diffusion is given as

$$\begin{cases} \frac{d}{dt}u_i(t) = f(u_i, v_i) + \epsilon \sum_{j=1}^{N} L_{ij}\, u_j \\ \frac{d}{dt}v_j(t) = g(u_i, v_i) + \sigma\epsilon \sum_{j=1}^{N} L_{ij}\, v_j \end{cases} \tag{4}$$

for $i = 1,2,\cdots,N$. Here, $\varepsilon$ $(= D_{act})$ and $\varepsilon\ \sigma$ $(= D_{inh})$ represent the diffusion mobility of the activator and the inhibitor, respectively.

The cellular automaton[41] is one approach for numerically simulating the reaction-diffusion model. In the cellular automaton, state, in addition to space, is discretized. For instance, the state has been discretized to two values, {1, 0}[42, 43]. This simple model is broadly known as the cellular automaton. Simulation based on the cellular automaton is executed step by step as a time evolution. The cellular automaton can form various patterns, which patterns have quite different behaviors, although they emerge under the same rules. The cellular automaton is described by the following time-evolution equation:

$$a_{i,j}^{(t+1)} = \phi\left[a_{i,j}^{(t)}, a_{i,j+1}^{(t)}, a_{i+1,j}^{(t)}, a_{i,j-1}^{(t)}, a_{i-1,j}^{(t)}\right] \tag{5}$$

for $i,j = 1,2,\cdots,N$. Here, $a_{i,j}^{(t+1)}$ represents a state on a node positioned at $(i, j)$ at time step $t$, and $\phi$ indicates a transition rule. The equation indicates that the next state depends on the states of neighboring nodes. It can be transformed into an equation like the reaction-diffusion model as follows:

$$u_i(t+1) = f\big(u_i(t)\big) + \sum_{\mathrm{j=1}}^{\mathrm{N}} \mathrm{L_{ij}}\, g\left(u_j(t)\right) \tag{6}$$

for $i,j = 1,2,\cdots,N$. This equation is also a time-evolution equation. To explain Eq. (6), a two-dimensional regular-grid network is considered as a topology. This regular-grid topology is a simple model. It is generally used to analyze spatial patterns of biological morphogenesis by using the reaction-diffusion model. In the regular-grid network, nodes are located at even intervals. Every node connects to the neighboring nodes, so the number of edges, called degree, is the same on all the nodes. As the number of dimensions is two, the degree is four (i.e., a node connecting with nodes above, below, to the left, and to the right). The patterns have been classified as four classes: stationary, periodic, chaotic, and complex[42, 43]. To characterize these classes, parameter $\lambda$ and mutual information have been introduced[40]. Parameter $\lambda$ is defined as $(K^N - n)/K^N$, where $K$ is the number of cell states, $N$ is degree, and $n$ is the number of rules that the state transfers to the quiescent state. Parameter $\lambda$ indicates the ratio of the total number of transition rules ($K^N$) and the number of rules that the state transfers to the quiescent state ($K^N - n$). The mutual information indicates the complexity of the pattern dynamics. These criteria are applied to the classification of the pattern. In class 1, $\lambda \approx 0$, and the mutual information is zero. The pattern in class 1 is stationary. In class 2, $\lambda < \lambda_c$, and the mutual information increases with $\lambda$. The pattern in class 2 is periodic. In class 3, $\lambda > \lambda_c$, and the mutual information decreases with $\lambda$. The pattern in class 3 is chaotic. In class 4, $\lambda = \lambda_c$, and the mutual information has a peak. The pattern in class 4 is complex. Formation of the patterns depend on $\lambda$. Thus, the various patterns emerge by using the reaction-diffusion model or the cellular automaton.

**Mutual Information.** To quantify patterns, mutual information as the complexity of a pattern has been introduced[40]. The entropy is defined as

$$H_i(\mathrm{t}) = -\sum_k P_i^k(t)\, log_k\, P_i^k(t) \tag{7}$$

where t indicates a time step, and *i* indicates the node label. $H_i(\mathrm{t})$ is the entropy at node *i* and at time step *t*. $P_i^k(T)$ is the occurrence probability of state *k*, and it is calculated for period *T*. State *k* is defined as that having the maximum number of particles. For example, when the number of states of a particle is two ($u$ and *v*), state *k* is also selected from $u$ and *v*. Namely, if $N_i^u > N_i^v$ at node *i*, $k = u$; otherwise, $k = v$.

The mutual information is obtained from the entropy as follows:

$$I_i(t, t+1) = H_i(t) + H_i(t+1) - H_i(t, t+1) \tag{8}$$

It follows that the mutual information of the entire network at time $t$ to $t+1$ is defined as

$$I_{ave}(t, t+1) = \frac{1}{N_{node}} \sum_i \left(I_i(t, t+1)\right) \tag{9}$$

The difference between the maximum mutual information and the minimum mutual information in the entire network is introduced to analyze the fluctuation in a pattern, and is defined as

$$I_{dif}(t, t+1) = \max_{i \in N_{node}} \left(I_i(t, t+1)\right) - \min_{i \in N_{node}} \left(I_i(t, t+1)\right) \tag{10}$$

**Amplitude and power.** To quantify patterns, the amplitude of the complexity of a pattern has been introduced[18]. The amplitude of an individual node *i* at time *t* is defined as

$$A_i(t) = \sqrt{[(u_i(t) - \langle u_i(t)\rangle)^2 + (v_i(t) - \langle v(t)\rangle)^2]} \tag{11}$$

where $\langle u_i(t)\rangle = \sum_i u_i(t)/N$ and $\langle v_i(t)\rangle = \sum_i v_i(t)/N$ are the mean quantities.

To compare patterns with a different number of particles ($r$), the amplitude is defined as

$$A_i(t) = \sqrt{[(N_i^u(t) - \langle N_i^u(t)\rangle)^2 + (N_i^v(t) - \langle N_i^v(t)\rangle)^2]} \tag{12}$$

where $\langle N_i^u(t)\rangle = \sum_i N_i^u(t)/N^u$ and $\langle N_i^v(t)\rangle = \sum_i N_i^v(t)/N^v$ are the mean quantities. The spatial-averaged amplitude is given as

$$\bar{A}(t) = \sqrt{\sum\nolimits_{\mathrm{i}} [(N_i^u(t) - \langle N_i^u(t)\rangle)^2 + (N_i^v(t) - \langle N_i^v(t)\rangle)^2]} \tag{13}$$

To quantify a temporal periodic pattern, the power spectrum of the amplitude, $A(f)$ , is introduced. $A(f)$ is obtained by applying a FFT (fast Fourier transform) to a time-series of $\bar{A}(t)$. Here, $f$ is a frequency, where $f = 1,2,\cdots,F$. When $f = 1$, $period = 0$ (i.e., a DC component), and when $f = 2,\cdots,F$, $period = T/(f-1)$ (i.e., AC components). The power of the AC component, $PW_{ac}$, is calculated by the following equation:

$$PW_{ac} = \frac{1}{F}\sqrt{\sum\nolimits_{\mathrm{f=2}}^{\mathrm{F}} |\mathrm{A(f)}|^2} \tag{14}$$

**Classification algorithm.** An algorithm for classifying the patterns by using the information-particle model is defined here. As a pseudo code, the algorithm is given as

```
TH_ave = 0.5;
TH_dif = 0.8;
TH_pw = 100;

if ((I_ave = 0) & (I_dif = 0))
        Class = 1;
    else if (((I_dif > TH_dif) & (PW_AC < TH_pw))
        Class = 2;
    else if ((I_ave > TH_ave) & (PW_AC > TH_pw))
        Class = 4;
    else
        Class = 3;
```

(15)

where $TH_{ave}$, $TH_{\mathrm{dif}}$, and $TH_{\mathrm{pw}}$ are thresholds for $I_{ave}$, $I_{\mathrm{dif}}$, and $PW_{AC}$, respectively. These values were obtained by the following numerical simulations for evaluating the information-particle model.

**Numerical simulations.** The information-particle model is given as Eq. (2), where $f(N_i^u, N_i^v)$ is given as

$$f(N_i^u, N_i^v) = \mathrm{a}N_i^u - \mathrm{b}(N_i^v)^2 - \mathrm{cMAX}(N_i^v - N_i^u + 1{,}0) \quad (16)$$

where $\mathrm{a} = 4, \mathrm{b} = 2, \mathrm{c} = 100,\ \varepsilon = 6,$ and $\sigma = 9/6$. The third term of Eq. (16) can be given simply as

$$f(N_i^u, N_i^v) = \mathrm{a}N_i^u - \mathrm{b}(N_i^v)^2 - \mathrm{c}N_i^v \quad (17)$$

Compared with Eq. (16), Eq. (17) tends to take many time steps to become stable. The patterns obtained by using Eq. (17) are the same as those obtained by using Eq. (16). Equation (16) was used to analyze the information-particle model. Other parameters and simulation conditions used in evaluating the model are listed as follows:

- Topology is a 2D regular-grid network.
  Number of nodes is 1024; *x* = 32 and *y* = 32.
- In the initial phase, all particles are located on a node at the center of the regular-grid network. The label of the center node is 528 (*x*=16, *y*=16).
- Number of states of a particle is two.
- Distance between nodes is 16.
- Time step ($\Delta t$) is 1.
- Velocity of a particle is 1.0 unit per time step; thus, the number of steps to traverse between nodes is 16.
- Stationary time at a node is two time steps.
- Number of time steps is two million.
- Time period for FFT analysis is 512 time steps.

## Author constitutions

J.M. designed the study and contributed to writing the paper. J.M. carried out the analysis and numerical simulation.

## Additional information

Authors declare no competing financial interests.

## Figures

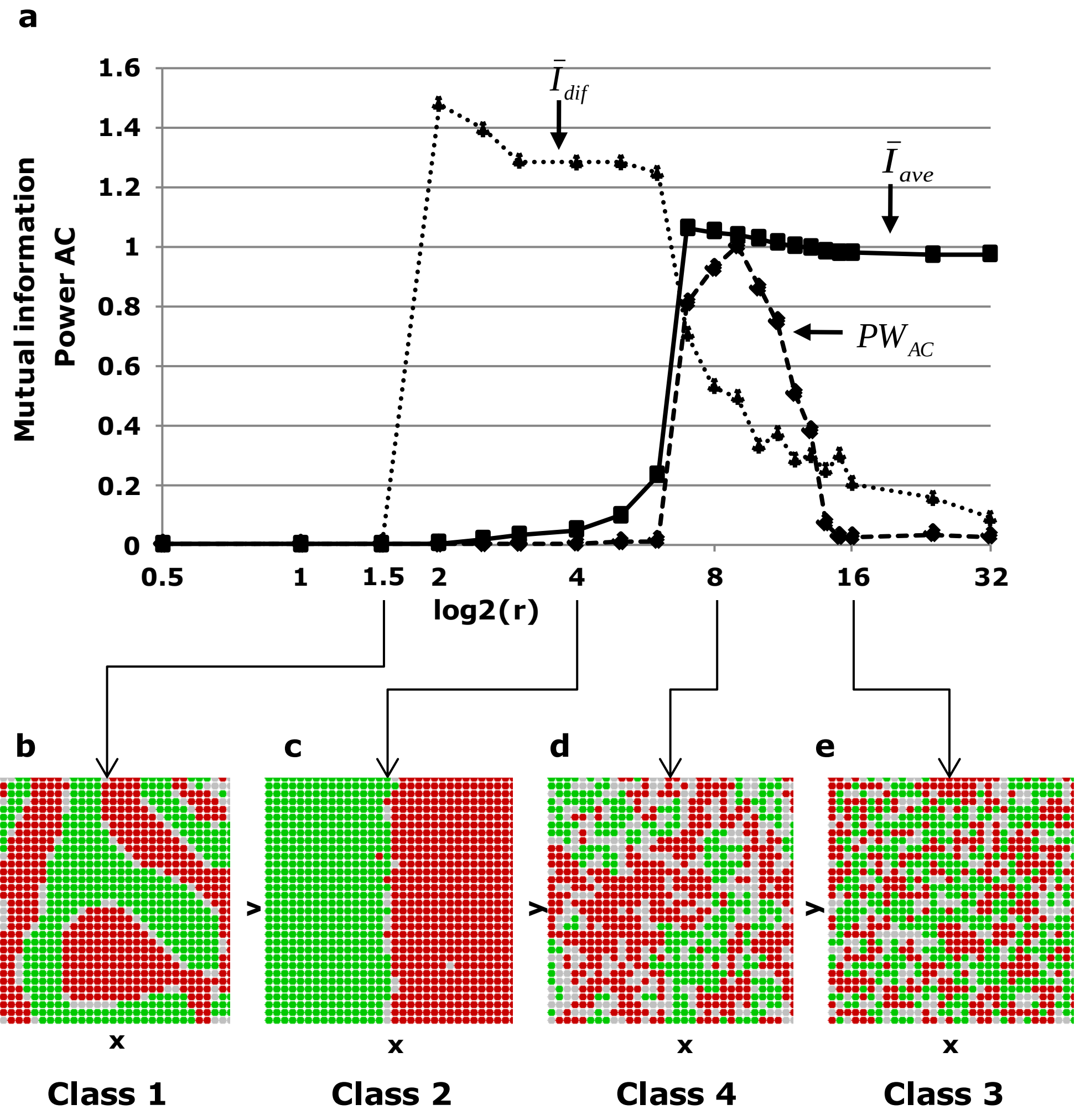

**Figure 1 | Pattern formations of the information-particle model.** In panel (a), mutual information ($I_{ave}$ and $I_{dif}$) and normalized power AC ($PW_{AC}$) are illustrated as a function of $N_{particle}/N_{node}$ (r). Panels (b), (c), (d), and (e) show patterns with $r$ of 1.5, 4.0, 8.0, and 16.0, respectively. There are two species, $u$ and $v$, in this simulation. If the number of particles in state $u$ is more than that in state $v$, the nodes are colored in green; otherwise, they are colored in red. The patterns are classified as "class 1", "class 2", "class 4", and "class 3", respectively.

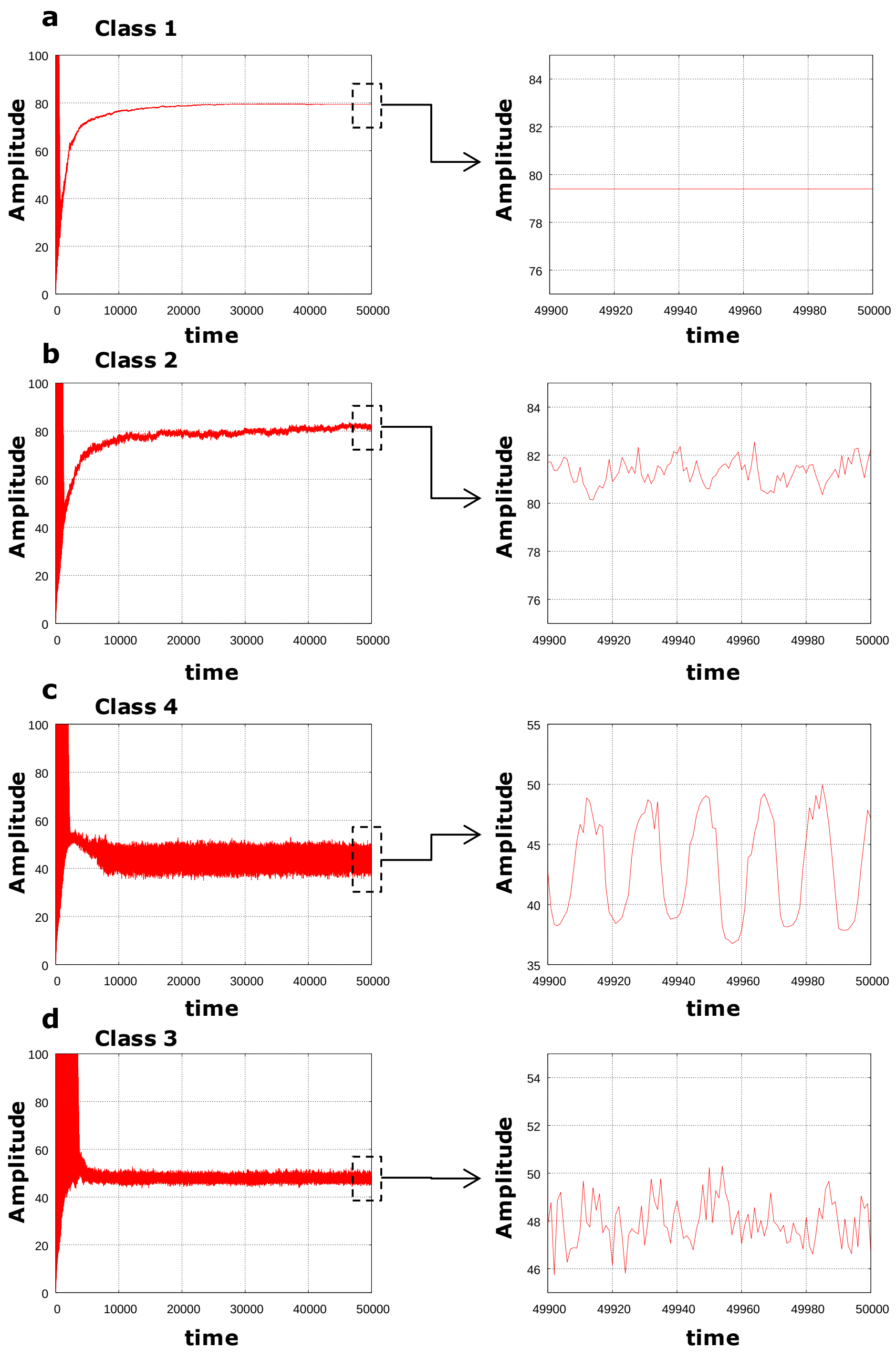

**Figure 2 | Temporal changes of amplitude in each class.** Left panels show the macroscopic behavior. Right panels show the zoomed temporal changes for $t$ = 49,900 to 50,000. The class and parameters are as follows: (a) class 1, $r$ = 1.5; (b) class 2, $r$ = 4; (c) class 4, $r$ = 8; and (d) class 3, $r$ = 16.

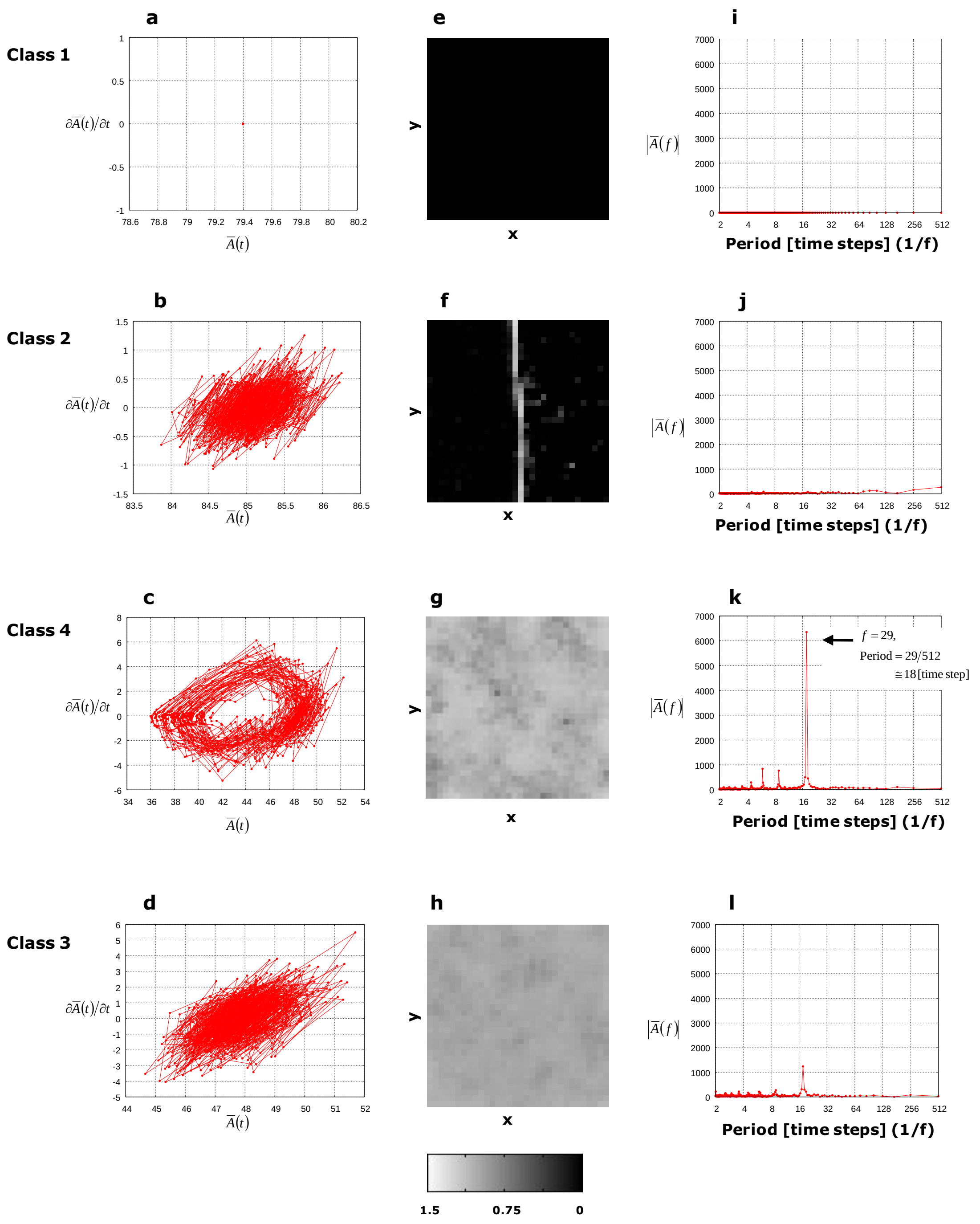

**Figure 3 | Analysis of patterns in each class.** Trajectory of amplitude. (a) to (d) show the trajectories of the amplitude. $\partial\bar{A}(t)/\partial t$ is defined as $\partial\bar{A}(t)/\partial t = \bar{A}(t+1) - \bar{A}(t)$. The period to display the trajectory is 1,000 steps after enough time (t = 1,990,000). (e) to (h) show the mutual information of spatial patterns. The dark color indicates that mutual information is low, and the light color indicates that mutual information is high. The map displays variation in space of the regular-grid network. *x* and *y* represent the axis of the network. (i) to (l) show power spectrum of the amplitude. The class and parameters are as follows: (a), (e), and (i): class 1, r = 1.5; (b), (f), and (j): class 2, r = 4; (c), (g), and (k): class 4, r = 8; (d), (h), and (l): class 3, r = 16.

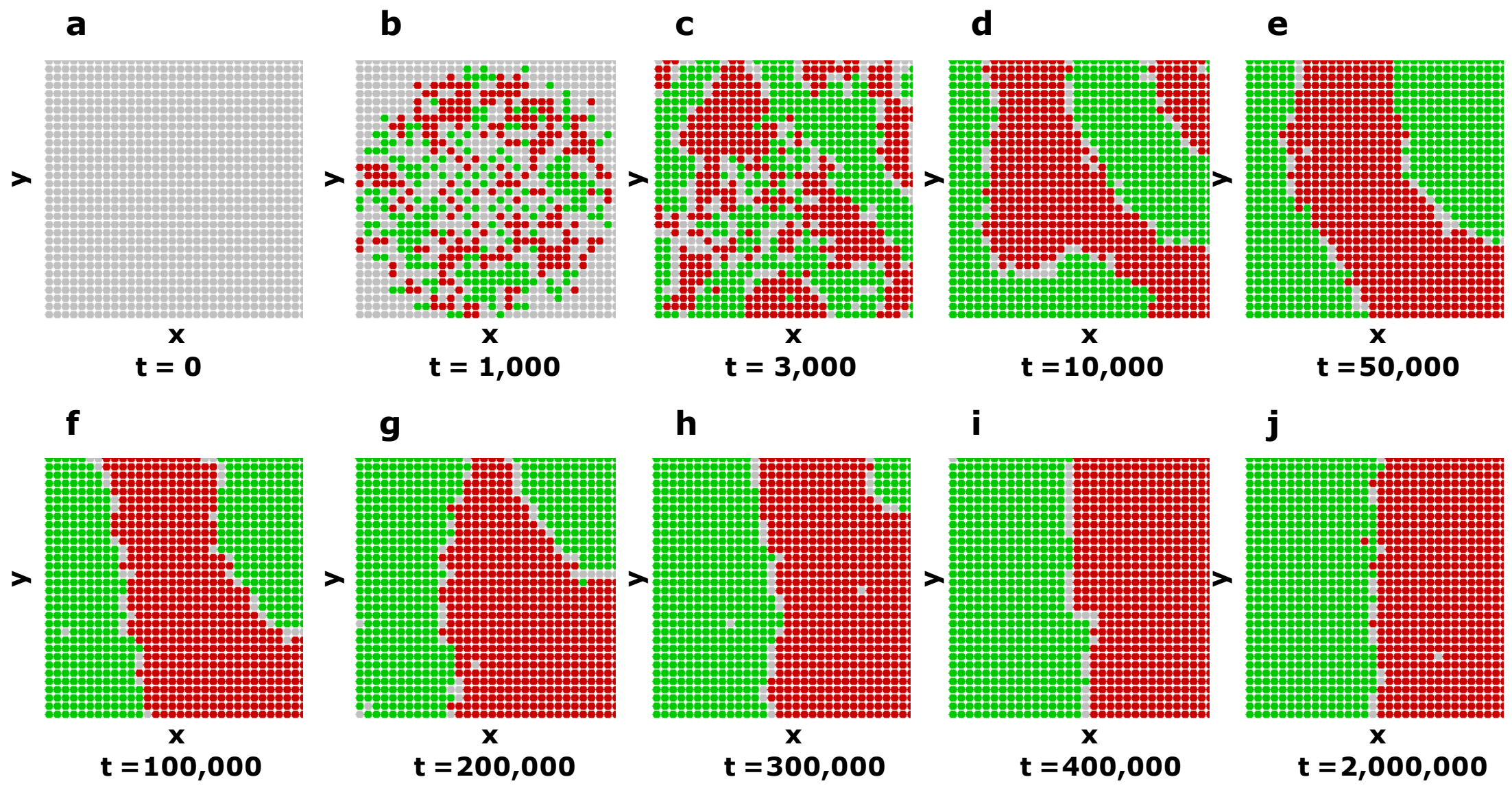

**Figure 4 | Dynamics of a pattern in class 2.** (a) to (j) show a snapshot at each time step. *x* and *y* represent the axis of the regular-grid network. There are two species, *u* and *v*, in this simulation. If the number of particles in state *u* is more than that in state *v*, the nodes are colored in green; otherwise, they are colored in red.

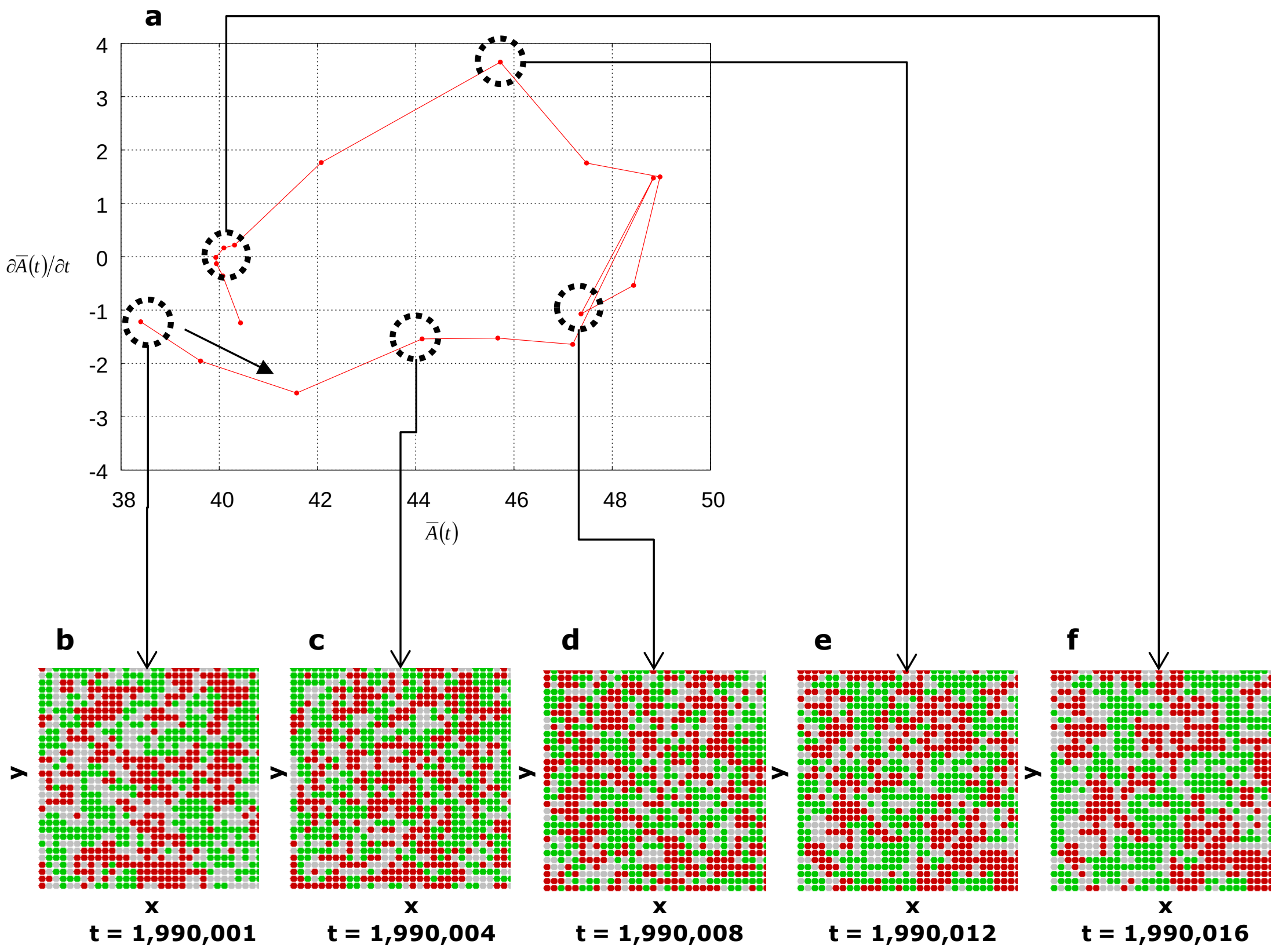

**Figure 5 | Dynamics of pattern in class 4.** (a) the trajectory of one period (18 time steps). 19 samples are plotted. Parameter *r* is 8. (b) – (f) show the patterns corresponding to each sample on the trajectory. (b) $t$ = 1,990,000; (c) $t$ = 1,990,004; (d) $t$ = 1,990,008; (e) $t$ = 1,990,012; and (f) $t$ = 1,990,016.

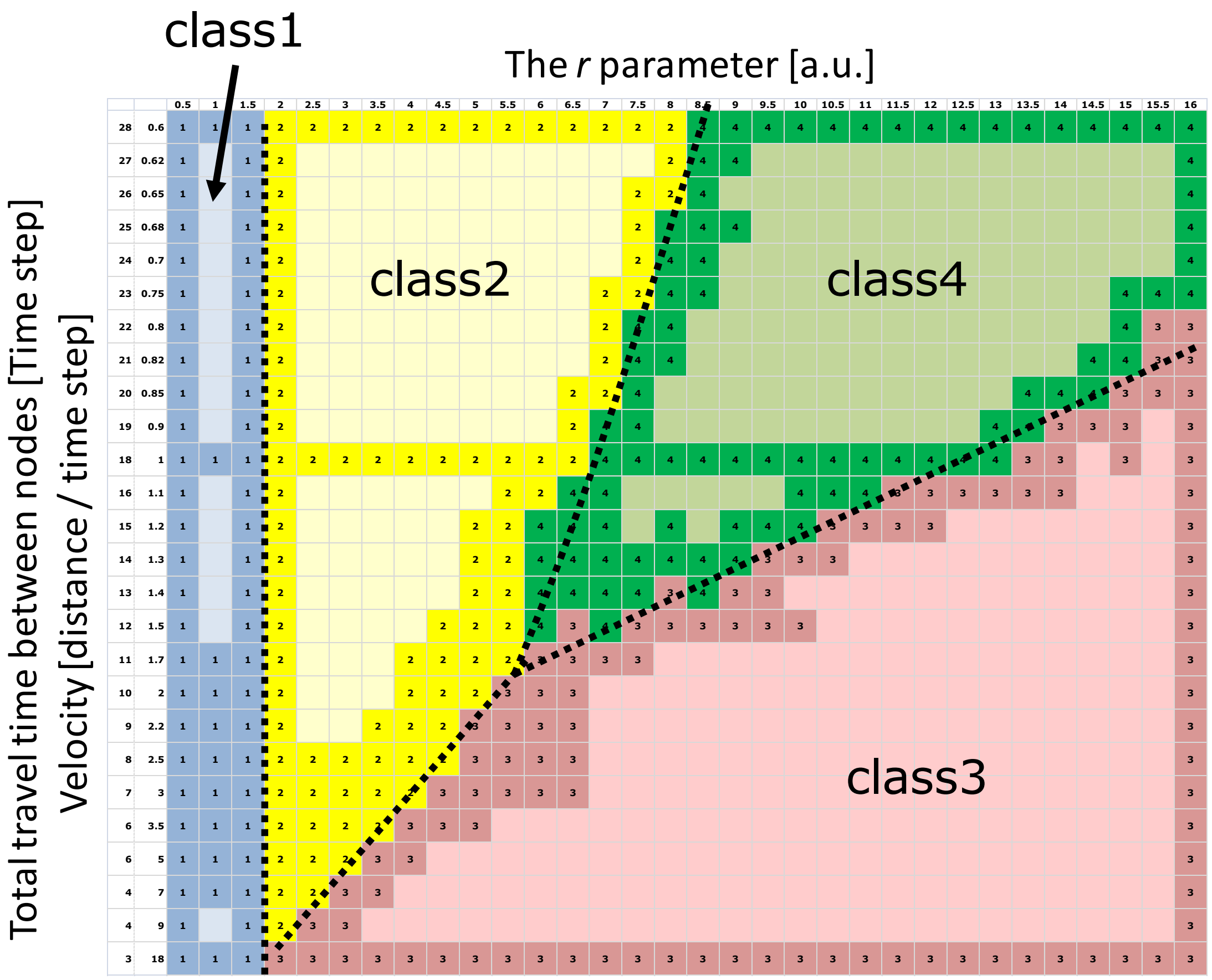

**Figure 6 | Class diagram of the patterns obtained by using the information-particle model.** Vertical axis shows travel time (velocity). Distance between nodes is 16. The stationary time on a node is two time steps; namely, the total travel time is 18 (i.e., 16 + 2) when the velocity is 1. The patterns shown in Fig. 1 are displayed on a row of travel time, i.e., 18 (velocity: 1). Horizontal axis shows parameter *r*. Each class is displayed in different colors. Simulated samples are indicated by a cell filled with a class number (classes 1 to 4). Classes of the other cells are estimated from the simulated samples. The class is obtained by the classification algorithm.